\documentclass[sigconf]{acmart}

\renewcommand\footnotetextcopyrightpermission[1]{}

\AtBeginDocument{%
  }

\begin{document}

\title{Cultural Newcomers Dining Across Borders: Need-Based Design Envision of Mixed Media Integration in MR for Foreign Menu Understanding and Ordering}


\author{Ying Zhang}
\affiliation{%
  \institution{Carnegie Mellon University}
  \city{Pittsburgh}
  \country{USA}}
\email{yingz5@andrew.cmu.edu}

\author{Daoxin Chen}
\affiliation{
  \institution{Indiana University Bloomington}
  \city{Bloomington}
  \country{USA}
}
\email{dc50@iu.edu}


\begin{abstract}
Cultural newcomers (CNs), including new immigrants and international students, often encounter cognitive barriers and social anxiety, exacerbated by unfamiliar cultural terminology in daily interactions. This research examines these challenges in the context of ordering in foreign restaurants. Current translation tools have significant limitations in their information delivery with current media presentation methods. This research investigates the challenges and needs of CNs in ordering scenarios in a foreign restaurant through interview sessions (N = 13) and explored their expectation of mixed media integration (Image, Video, 3D Model) through a participatory design session that featured an immersive restaurant experience to support brainstorming. Based on qualitative analysis of participants' needs and expectations, the mixed media ordering assistant is conceptualized across 4 key dimensions: Key Features, User interaction, Media hierarchy, and Information presentation, with the objective of alleviating cultural barrier, linguistic barrier, cognitive load and improving the dining experience for CNs.
\end{abstract}




\maketitle
\thispagestyle{empty}
\pagestyle{empty}

\section{Introduction}
In an increasingly globalized world, cultural newcomers (CNs), including international students and new immigrants, face numerous challenges as they navigate unfamiliar environments. These individuals are frequently exposed to specialized terminology and cultural practices that are foreign to them in their daily lives. One of the most common scenarios where this cultural and linguistic barrier becomes particularly evident is in restaurant settings, specifically during the ordering process.

The United States, as a major destination for international students and immigrants, hosts a substantial number of cultural newcomers each year. According to the Institute of International Education, there were more than 1 million international students in the United States during the academic year 2022-2023\cite{Statista2023}. Most CNs who lack prior experience living abroad are often confronted with significant cultural and linguistic barriers, particularly in the context of dining. Cuisine, as a crucial element of cultural identity, varies widely across cultures and can be a source of both intrigue and confusion for CNs.   

Many restaurant menus in the US feature dishes with names or descriptions that include foreign language terms or are derived from culinary traditions other than the dominant culture. According to the cognitive load theory, individuals processing new and multiple types of information can experience decreased performance and increased mental effort.\cite{VanMerrienboer2005} The language barrier in dining situations extends beyond mere translation issues. Many culinary terms, dish names, and ingredients are deeply rooted in specific cultural contexts, making them difficult to understand without prior knowledge or experience. This linguistic-complex ordering experience in a foreign country can significantly increase the cognitive load for CNs, making the ordering process more challenging and potentially hindering their cultural adaptation.

Although there is existing research on language support for cultural newcomers, there is a notable gap in addressing the specific needs related to specialized terminology and cultural practices in everyday contexts, particularly in dining situations. This study aims to bridge this gap by exploring mixed media technology solutions to support the ordering experience focusing on the following objectives: a) To identify and analyze the ordering needs and challenges faced by CNs in restaurant settings. b) To explore the potential integration of mixed media technologies in Mixed Reality (MR) to facilitate the ordering experience for cultural newcomers. c) To assess the acceptability and perceived effectiveness of mixed media methods in MR as supporting tools for menu understanding and order decision-making. By addressing the specific challenges related to menu comprehension and food ordering, this research can contribute to the development of more effective and targeted support systems for CNs and facilitate easier navigation of cultural and linguistic barriers in common social interactions, thereby enhancing the integration experience for CNs. 

To achieve these objectives, the study will address the following research questions:

Q1: What are cultural newcomers' current methods of learning about cuisine, and what are the limitations of these methods?

Q2: What specific needs do cultural newcomers have in understanding foreign menus and making ordering decisions?

Q3: To what extent are mixed media methods in MR accepted as "supportive" for the ordering experience (menu understanding and order decision making)?

Q4: What are users' expectations for integrating mixed media in MR to facilitate the ordering experience?

By answering these questions, this study aims to contribute to the expanding research on cultural adaptation and technology-assisted learning. The findings will offer valuable insights for the development of digital language support tools, particularly in designing user-centered mixed reality (MR) applications that integrate mixed media to facilitate the understanding of culturally specific terms. 

\section{Related work}
\subsection{Challenges in Understanding Cuisine Names from Different Cultures}

Food is a powerful medium for cultural expression and identity. Ahn (2011)\cite{Ahn2011} demonstrated that culinary terms and flavor combinations are deeply rooted in cultural contexts, with the names and compositions of dishes encoding traditions that are largely opaque to outsiders without prior cultural knowledge. This opacity is compounded in cosmopolitan dining environments, where menus frequently draw from multiple culinary traditions and languages simultaneously. Peng (2015)\cite{Peng2015} examined the translation of dish names, revealing the inherent difficulty of conveying cultural nuance and culinary convention through direct linguistic transfer alone. 

Menus in cosmopolitan areas often feature a mix of languages, which can pose a significant linguistic challenge to diners who may not be fluent in the local language or the language from which the dish originates. Harmon(2019)\cite{HarmonJones2019} theory of cognitive dissonance mentioned that when an individual has conflicting cognitions, there could be a mental discomfort experience. Frank (2021)\cite{Frank2021} provides information on the cognitive challenges of multilingual sentence processing using computational models. In the context of foreign menus, especially multilingual menus, diners can experience more cognitive challenges when they encounter unfamiliar languages, leading to confusion and hesitation in making dining decisions.

Sweller’s cognitive load theory\cite{VanMerrienboer2005} suggests that increased mental effort is required for decision making when dealing with unfamiliar and complex information, such as names of foreign cuisine. This increased cognitive load can affect the ability of the diner to process information effectively, leading to decision fatigue and reduced dining satisfaction. Although the research does not focus directly on CNs, it elucidates potential challenges that they may encounter during the ordering process. 

\subsection{Informational Media and Cognitive Load}

\begin{figure}[h]
  \centering
  \includegraphics[width=0.5\linewidth]{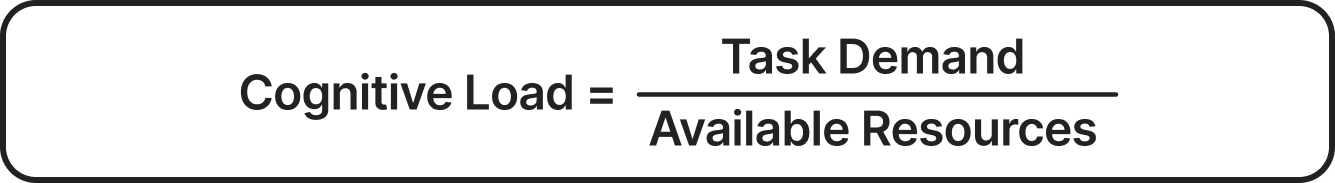}
  \caption{Cognitive Load Theory (Frederick Reif)}
  \label{fig:cognitive-load}
  \Description{}
\end{figure}

\begin{figure}[h]
  \centering
  \includegraphics[width=0.6\linewidth]{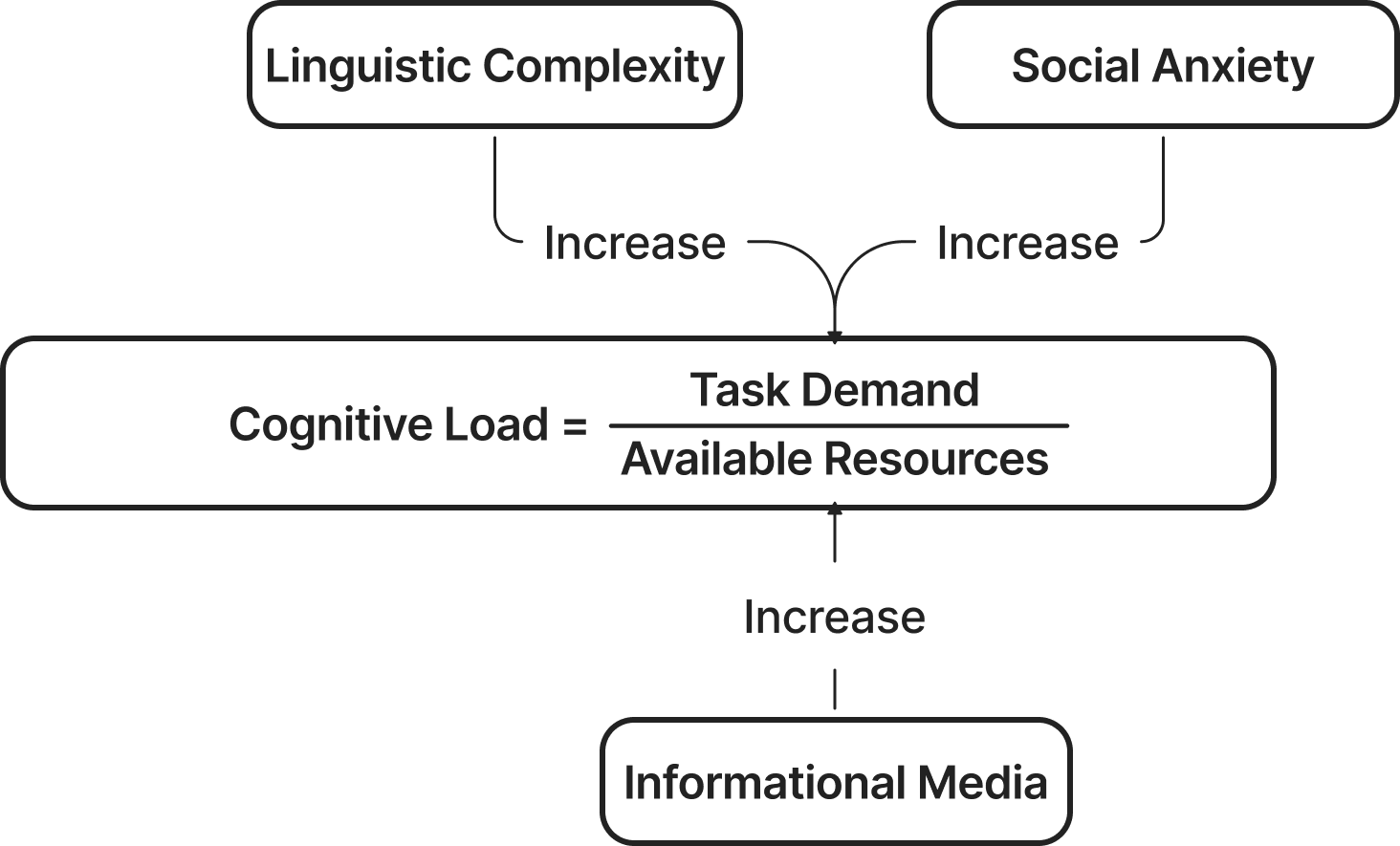}
  \caption{Informational Media and CNs' Challenges impact on Cognitive Load }
  \label{fig:CLT_N}
  \Description{}
\end{figure}

Sweller and Chandler (1991)\cite{SwellerChandler1991} think that the learning of individual elements is the simplest, and the cognitive load increases when learners also need to know the interactivity between different elements. This interactivity between different elements can be triggered by the intrinsic cognitive load and extraneous cognitive load (Sweller \& Chandler, 1991).\cite{SwellerChandler1991} Intrinsic cognitive load is associated with the inherent complexity of the content, while extraneous cognitive load is related to the way the information is formatted and presented, which can be influenced by the use of mixed media (Sweller and Chandler, 1991)\cite{SwellerChandler1991} 
.
As illustrated in Figure~\ref{fig:cognitive-load}, Reif (2010)\cite{reif2010applying} presents the Cognitive Load Theory through a concise formula. Considering the relationship between cognitive load and traditional media, including texts, images with noninteractive stimuli, Mayer (2001)\cite{Mayer2001} put forward a theory of mixed media learning, stating that learners can better create mental representations of external information when it is presented through multiple sensory pathways. Thus, the impact of informational media on cognitive load is illustrated in Figure~\ref{fig:CLT_N}. With the population of interactive mixed media including video games and virtual reality, more researchers focused on the influence of mixed media interaction for cognition, particularly hands-on learning (Zheng et.al., 2009)\cite{Zheng2009}. For example, Engelkamp (1998)\cite{Engelkamp1998} thinks that observing actions use encoding processes from performed actions, and therefore argues that learning through manipulation, should be encoded differently from learning through verbal and visual inputs. Dining in an unfamiliar cultural context can often lead to social anxiety. Hofmann (2010)\cite{Hofmann2010} defines social anxiety disorder as an excessive fear of violating social norms, a concept that is closely intertwined with the fear of embarrassment. Huk and Ludwigs (2009)\cite{Huk2009} demonstrate that cognitive and affective supports, when combined, significantly enhance learning outcomes in an economics simulation among social science students.

\subsection{Technology Intervention and Supportive tools for Culture Newcomers}
Several prior efforts have applied extended reality (XR) to food-related contexts. Ajune W.\cite{IsmailFadzli2025} explored the food ordering system in XR with a focus on presenting menu options. Velasco et al. (2021)\cite{Velasco2021} introduce a model for creating “impossible” food experiences using XR, which includes elements that defy physical laws and incorporate fantasy. 
The CoDine system by Wei et al. (2011)\cite{UbiComp2011} connects people in different locations through shared dining activities, using gesture-based interactions and multisensory communication.
Current explorations on XR dining do not focus on facilitating the ordering experience itself. 

This systematic review by Zain (2023)\cite{Zain2023} examines the use of augmented reality (AR) in language teaching and learning. Sabie et al. (2023)\cite{Sabie2023} discuss the use of AR for intercultural heritage exchange, emphasizing its role in enhancing cultural understanding and engagement. Current research focuses more on formal cultural education but does not extend to casual, everyday scenarios such as dining. There is a  gap in studies specifically examining how XR can be used to bridge cultural gaps for CNs through casual and daily intervention. 

The current mixed media integration explorations in XR are more about cultural introduction and learning in a more formal scenario. There is a gap in the current technology solution to integrate mixed media into solutions for more casual and daily scenarios to facilitate CNs, such as the ordering experience.

\subsection{The Underaddressed Daily Needs of Cultural Newcomers in HCI}
Human-food interaction, as a subfield of HCI, addresses the ways in which digital technology can enrich the relationship between people and food across dimensions of commensality, playful engagement, inclusive design, healthy eating, and sustainable consumption (Deng et al., 2023)\cite{Deng2023}. Within this space, XR has been applied to restaurant contexts with promising results. Seetharam et. al. (2023)\cite{Seetharam2023} designed an AR-based web solution enabling diners to visualize and compare dishes prior to ordering, incorporating a chatbot to deliver personalized recommendations and reduce food waste. Ismail \& Fadzli (2024)\cite{IsmailFadzli2024} similarly developed an interactive AR application bridging the physical and digital dimensions of the dining experience.

Despite this activity, the application of XR to the specific challenges of cultural newcomers — particularly the cognitive and social difficulties involved in deciphering foreign menus — has received little systematic attention. Existing tools tend to address general users rather than those navigating a fundamental linguistic and cultural gap. This study addresses that gap directly by investigating how mixed media integration within a mixed reality environment can be designed to support cultural newcomers' menu comprehension and ordering decisions, reducing cognitive load while acknowledging the social and affective dimensions of cross-cultural dining.

\section{Methodology}
Prior to the interview,  we designed the questionnaire as a preparation stage to find out the current stages of  using digital tools to solve the problems in the process of  food ordering. We designed the questionnaire into two sections, namely the basic demographic information, and the frequency of using the digital tools when they look through the menu and order food in restaurants and their satisfaction about these tools. Through using these questionnaires, we aimed to have an overall expectation of the degree that newcomers rely on the digital tools and the limitations of the current tools, which would be a basis for the design of the following interview. Through distributing questionnaires among different social media platforms, we collected 50 qualified data from international students and newcomers of the host countries.

\begin{table*}
  \caption{Descriptive Statistics for Length of Stay and Need for Menu Explanations}
  \label{tab:Statistic}
  \begin{tabular}{llcc}
    \toprule
    Characteristics & Sub Level & Frequency & Percentage\\
    \midrule
    \texttt{Duration in U.S.} & {Less than 1 year} & 14 & 28\% \\
    \texttt{} & {1 - 3 years} & 8 & 16\% \\
    \texttt{} & {3 - 5 years} & 11 & 22\% \\
    \texttt{} & {More than 5 years} & 17 & 34\% \\
    \midrule
    \texttt{Menu Explanation Needs} & {Never} & 3 & 6\% \\
    \texttt{} & {Rarely} & 9 & 18\% \\
    \texttt{} & {Sometimes} & 22 & 44\% \\
    \texttt{} & {Often} & 13 & 26\% \\
    \texttt{} & {Always} & 3 & 6\% \\
    \bottomrule
  \end{tabular}
\end{table*}

\begin{table*}
\centering
\caption{Descriptive Statistics for the Frequency of Tool Usage}
\label{tab:tool_usage}
\begin{tabular}{lcccc}
\toprule
Tool & Average & Count \\
\midrule
Review Apps (Google Map) & 3.00 & 38 \\
Google Search & 3.08 & 37 \\
Translation Softwares & 2.81 & 36 \\
Social Media & 2.91 & 35 \\
Restaurant Website & 2.29 & 35 \\
\bottomrule
\end{tabular}
\\[0.5ex]
{\small Note: The usage frequency was measured on a 5-point Likert scale: 1 (Never) to 5 (Always).}
\end{table*}

\begin{table*}
\centering
\caption{Satisfaction Ratings for Digital Tools Used}
\label{tab:satisfaction_ratings}
\begin{tabular}{lcc}
\toprule
Tool & Average Satisfaction & Count \\
\midrule
Google Map (Reviews) & 3.48 & 29 \\
Google Search & 3.67 & 30 \\
Translation Softwares & 3.21 & 28 \\
Social Media & 3.85 & 26 \\
Restaurant Website & 3.14 & 21 \\
\bottomrule
\end{tabular}
\\[0.5ex]
{\small Note: This satisfaction was measured on a 5-point Likert scale: 1 (Very Dissatisfied) to 5 (Very Satisfied).}
\end{table*}

\subsection{Study Design}

\begin{figure*}[h]
  \centering
  \includegraphics[width=\linewidth]{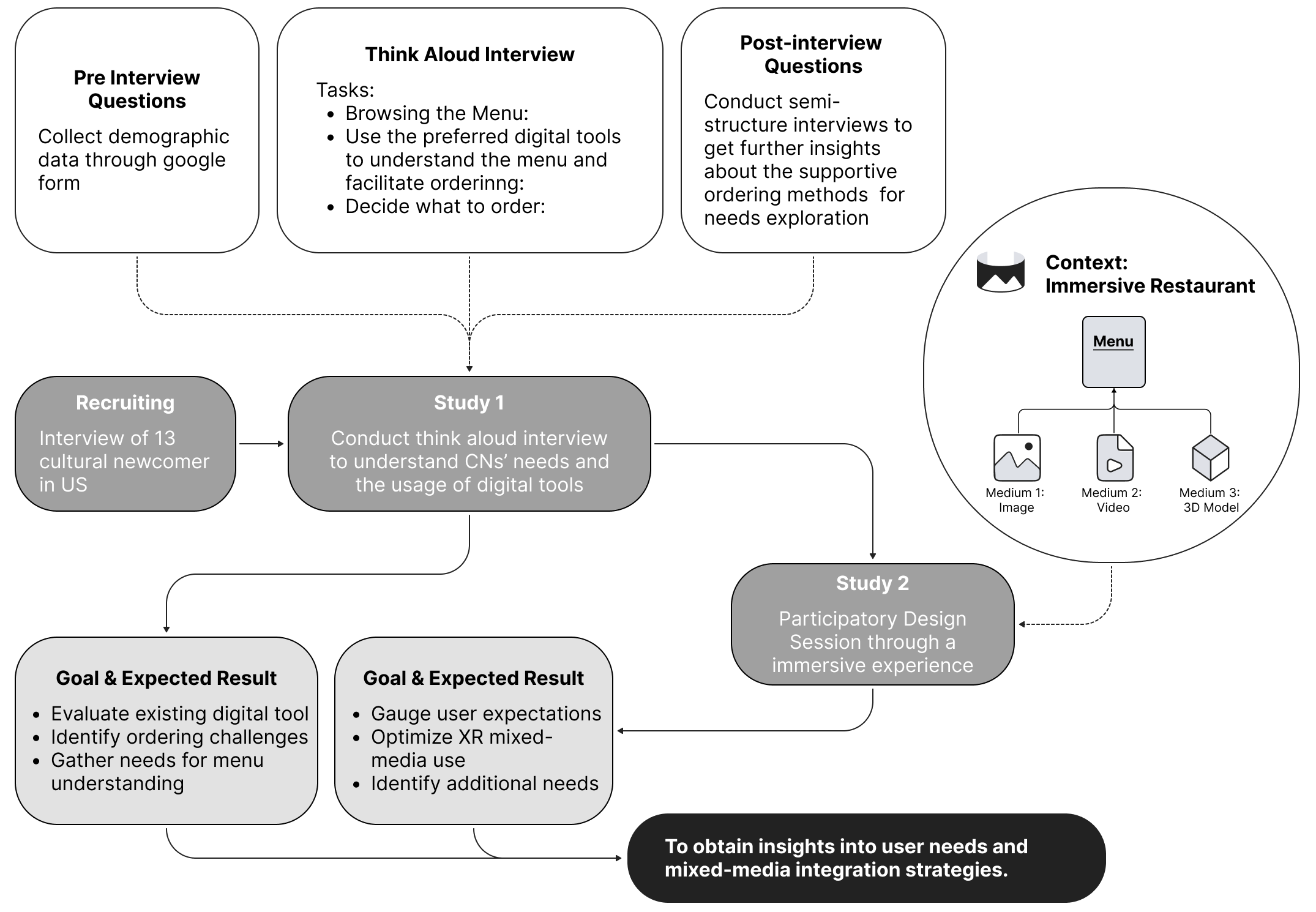}
  \caption{Study Design Diagram}
  \label{fig:study design}
  \Description{}
\end{figure*}

    After the questionnaire, to get in-depth understanding of CNs’s needs, and explore the potential of mixed media integration in mixed reality to facilitate the ordering experience. We employ a multi-phase mixed methods approach and qualitative and experimental strategies to explore the informational needs and preferences of CNs in menu comprehension and order decision. The first phase is to conduct a thought-aloud interview to identify the initial challenges and informational needs encountered by CNs when ordering food in a foreign restaurant. The second phase is an immersive experiment to explore the effectiveness of mixed media integrations in an MR environment in improving the ordering experience and participants' expectations for mixed media integration in MR. 

    We recruited participants (N = 13), including international students and new immigrants in the United States with varying lengths of stay and diverse backgrounds. The primary criterion for inclusion was that participants had experienced difficulty understanding foreign menus in restaurant settings. We aim for a diverse sample in terms of cultural background, language proficiency, and familiarity with XR technologies to ensure a wide range of perspectives.  

    All sessions were video and audio recorded with participant consent. Screen recordings were also captured when participants used digital tools on the provided mobile device. This multi-modal data collection allowed for a comprehensive analysis of participants' verbal reports, non-verbal cues, and digital tool usage.
\subsection{Think aloud interviews}
To gain in-depth insight into the challenges faced by cultural newcomers (CNs) when navigating foreign menus and their current strategies for interpreting menus, we conducted think-aloud interviews. This method allowed us to observe participants' thought processes and behaviors in real time as they interacted with unfamiliar menus. The think-aloud interview consists of pre-interview questions, think-aloud interviews, and post-interview questions. We began with the pre-interview questions to gather demographic information and initial insights into participants' experiences with foreign menus. Followed by the think-aloud interviews, the participants were presented with unfamiliar menus and asked to verbalize their thoughts as they navigated them. 

The ordering experience consists of two stages: 1) understanding the menu, 2) ordering decision making. For the rest of the paper, the discussion will focus on the two stages, when it comes to 'ordering experience', it represents a simplified representation of the two stages. And to get insights into the specific needs for different stages, the think-aloud interview included three main tasks:
\begin{itemize}
\item {\texttt{Menu browsing}}: Participants were given 3 minutes to browse an unfamiliar menu while thinking aloud.
\item {\texttt{Menu exploration}}: Participants used their preferred digital tools to understand unfamiliar menu items.
\item {\texttt{Order decision}}: The participants made hypothetical food orders based on their understanding of the menu.
\end{itemize}
After the think-aloud interview, the participants were asked a series of follow-up questions to get additional insights about their overall comments on the digital tools and their uncovered needs. 

\subsection{Participatory Design: Immersive Experiment}

\begin{figure*}[h]
  \centering
  \includegraphics[width=\linewidth]{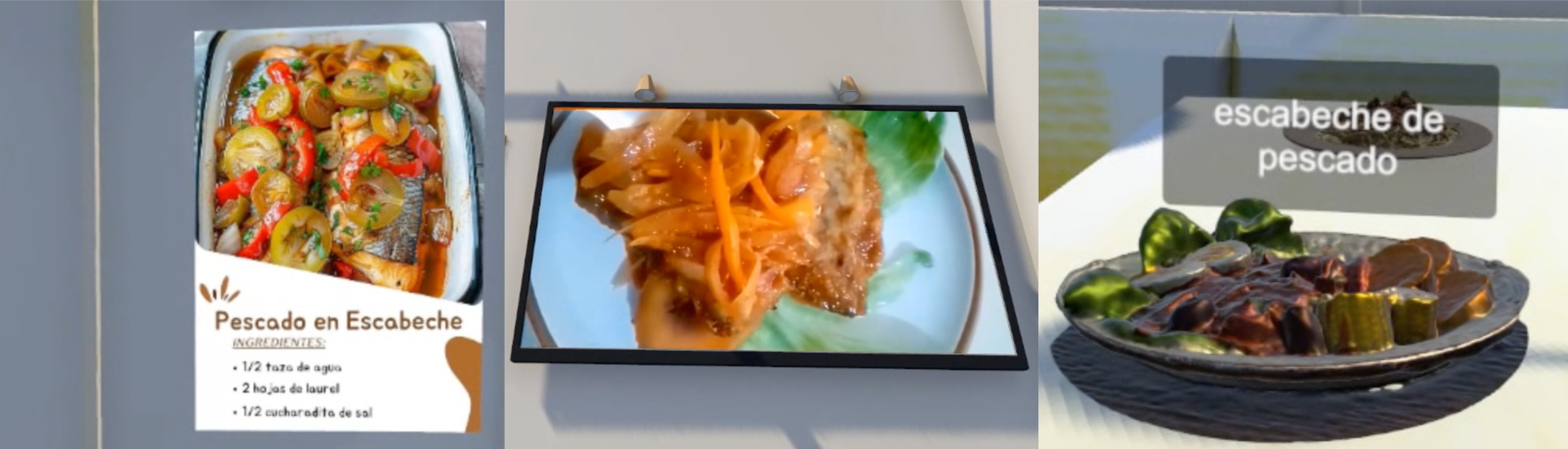}
  \caption{Media integrated in the immersive experiment, from left to right: Image, Video, 3D Model }
  \Description{screenshot from the experiment.}
  \label{fig: Screenshot VR}
\end{figure*}

In this phase, we developed an immersive restaurant environment integrated with multiple media including image, video, and 3d model in Unity and built it into Oculus Quest headsets. Participants were invited to experience this immersive restaurant and interact with various integrated media. During this session, participants were guided to get familiarity with XR interactions and explore different media in the immersive experience. 

\textbf{Immersive environment setting and media integration:}
To provide an intuitive understanding for the participants to understand the perception feeling of the media in XR environment,  we incorporated three primary types of media to facilitate presentation of the cuisine.
\begin{itemize}
\item {\texttt{Images}}: High-resolution photographs of dishes
\item {\texttt{Videos}}: Video clips demonstrating food preparation and plating.
\item {\texttt{3D Models}}: Static models of dishes placed on the table.
\end{itemize}

\textbf{Semi-structured interview:} During the immersive experience, we conducted semi-structured interviews to obtain user feedback on different media and their expectations of the mixed media redesign and integration to facilitate the ordering experience. During the semi-structured interview, we focus on gaining insights about:
\begin{itemize}
\item Willingness to use and perceived helpfulness of each media format
\item Preferences for different media types
\item Ideas of interaction methods with mixed media
\item Ideas of integration multiple media as an ordering assistive tool
\end{itemize}

\begin{figure*}[h]
  \centering
  \includegraphics[width=\linewidth]{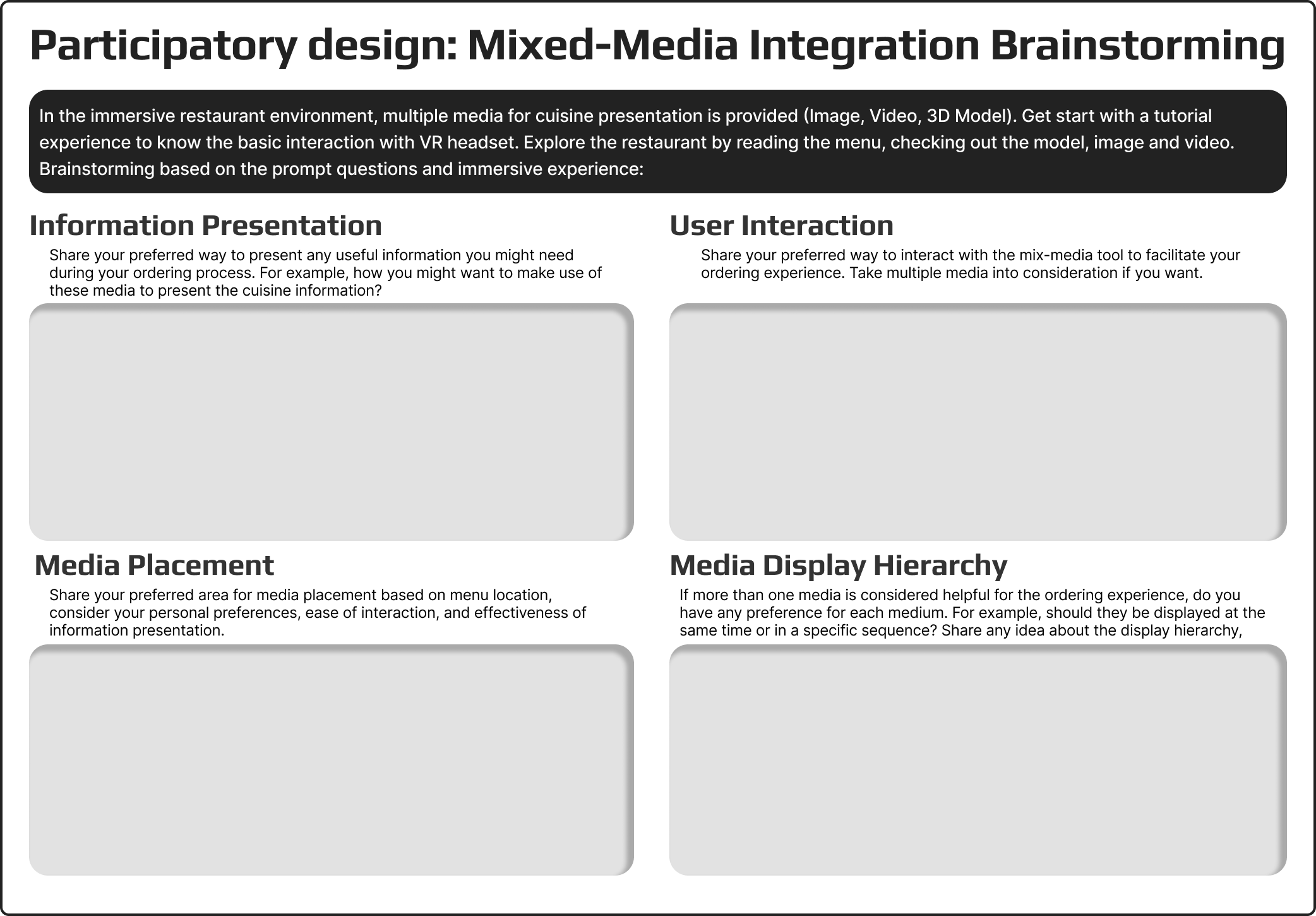}
  \caption{Participatory Design Brainstorming Board}
  \Description{}
  \label{fig: brainstorming board}
\end{figure*}

\textbf{Participatory Design Session:} In the immersive restaurant environment, multiple media are provided for presentation of the cuisine (image, video, 3D model). Figure ~\ref{fig: brainstorming board} is a participatory design board for participants to brainstorm and share design ideas. Participants are guided to start with a tutorial experience to learn the basic interaction with VR headset. After exploring the restaurant by reading the menu, checking out the model, image, and video, participants are invited to brainstorm based on prompt questions and immersive experience to share any design ideas about:
\begin{itemize}
\item \textbf{Information Presentation:} Verify the need of information to facilitate ordering experience, and identify participants' preferred way to present their needed information through different media
\item \textbf{User Interaction:} Identify the expected interaction of the participants with the mixed media tool. 
\item \textbf{Media Placement:} Identify the preferred placement area based on different interaction assumptions with various media.
\item \textbf{Media Display Hierarchy:} Identify the participants' preferred display hierarchy of each medium that might be related to their necessity and convenience.
\end{itemize}

\subsection{Data Collection and Analysis}

\begin{table*}
\centering
\caption{Demographic information of CNs}
\label{tab:demographic}
\begin{tabular}{lccccc}
\toprule
No. & Gender & Age & Duration in U.S. & Ordering Challenge Level & Menu Difficulty Emotions \\
\midrule
P1 & F & 19 - 25 & 0 - 1 year & Very challenging & Anxiety, Curiosity \\
P2 & F & 19 - 25 & 0 - 1 year & Extremely challenging & Anxiety, Embarrassment \\
P3 & M & 19 - 25 & 1 - 3 years & Moderately challenging & Embarrassment, Curiosity \\
P4 & M & 26 - 35 & 3 - 5 years & Moderately challenging & Curiosity \\
P5 & F & 26 - 35 & 3 - 5 years & Moderately challenging & Frustration, Embarrassment \\
P6 & M & 19 - 25 & Above 5 years & Moderately challenging & Anxiety, Embarrassment \\
P7 & F & 26 - 35 & 1 - 3 years & Moderately challenging & Curiosity \\
P8 & F & 19 - 25 & 3 - 5 years & Slightly challenging & Frustration, Curiosity \\
P9 & M & 19 - 25 & 0 - 1 year & Moderately challenging & Indifference \\
P10 & F & 26 - 35 & 0 - 1 year & Very challenging & Embarrassment, Indifference \\
P11 & M & 26 - 35 & 0 - 1 year & Moderately challenging & Frustration, Anxiety, Embarrassment\\
P12 & F & 19 - 25 & Above 5 years & Slightly challenging & Curiosity \\
P13 & Non-binary & 26 - 35 & Above 5 years & Moderately challenging & Curiosity, Indifference \\
\bottomrule
\end{tabular}
\end{table*}

Data was collected through multiple methods:
\begin{itemize}
    \item Questionnaire responses to collect demographic information and initial insights into participants' experiences with foreign menus.
    \item Audio and video recordings of think-aloud sessions and interviews
    \item Screen Recordings of interaction with the digital tools
    \item Researcher observations and notes
\end{itemize}
We employ a mixed-method analysis approach.

\begin{itemize}
    \item Quantitative analysis of Likert-scale ratings for media preferences and helpfulness
    \item Qualitative thematic analysis of interview transcripts and think-aloud session data and semi-structured interview section data.
\end{itemize}

\section{Findings}

\subsection{Current Available Tools}

Participants universally relied on a combination of digital tools to compensate for the limitations
of any single application. Rather than finding one tool sufficient, all 13 participants described
switching among multiple tools within a single ordering episode — a behaviour that itself signals
a fundamental gap in current solutions.

Table~\ref{tab:tools_consolidated} consolidates the tool landscape into a single reference,
extending the descriptive categories from the questionnaire phase with the key limitation
identified in the think-aloud sessions for each tool type.

\begin{table*}
\centering
\caption{Digital Tools Used by Cultural Newcomers: Mode, Media Output, and Key Limitation}
\label{tab:tools_consolidated}

\begin{tabular}{p{1.8cm} p{3.6cm} p{1.2cm} p{1.2cm} p{5.2cm}}
\toprule
\textbf{Tool} &
\textbf{Primary Use} &
\textbf{Input} &
\textbf{Output Media} &
\textbf{Key Limitation (think-aloud)} \\
\midrule

AI Tools (e.g., ChatGPT) &
Summarised explanations; conversational recommendations via chat &
Text / voice &
Text only &
No visual output; users must switch tools for images. Cannot show dish appearance directly. \\

Search Engine (e.g., Google) &
Broad information retrieval across web content &
Text &
Image, text, video &
Images are not dish-specific; portion size unclear; weak mapping between name and real dish. \\

Review Apps (e.g., Google Maps, Yelp) &
Peer reviews and restaurant food photos &
Text &
Image, text &
Sparse coverage for small restaurants; dish names hard to trace back to menus. \\

Photo-based Translation App &
Real-time menu translation via camera &
Camera / photo &
Text overlay &
Dish-name translation unreliable; ingredient translation more accurate than titles. \\

Text-based Translation App &
Manual translation of dish terms &
Typed text &
Text only &
Slow and inefficient; no visual context; weak semantic accuracy for dish names. \\

\bottomrule
\end{tabular}
\end{table*}

Two patterns emerged consistently. First, \textit{information fragmentation}: each tool delivers
a single modality, forcing participants to triangulate across applications to assemble a
sufficiently complete picture of a dish. A typical sequence involved photo-translation to obtain
ingredient terms, followed by a search-engine image lookup, followed by social media to verify
restaurant-specific presentation. Second, \textit{name-continuity failure}: once a dish was
identified through one tool, relocating its name on the physical menu proved difficult, creating
a secondary burden that sometimes required asking waitstaff for assistance — the very
interaction participants most wished to avoid.

\subsection{Needs and Challenges in Foreign Menu Ordering}

\subsubsection{Informational Needs}

Eleven of 13 participants reported needing information about dish \textit{size}, and more than
half needed information about \textit{flavour} (e.g., spiciness level) — neither of which is
reliably conveyed by existing tools or standard menu formats. Ingredients were prioritised over
dish names: when names were opaque, participants redirected attention to descriptions and
ingredient lists as the primary basis for decisions. Five participants additionally sought
information about cooking methods, utensils, and seasoning. Table~\ref{tab:needs} summarises
the full need taxonomy.

\begin{table*}
\centering
\caption{Taxonomy of Ordering Needs Identified Through Think-Aloud Interviews (N=13)}
\label{tab:needs}

\begin{tabular}{p{4cm} p{8cm}}
\toprule
\textbf{Category} & \textbf{Specific Needs} \\
\midrule

Cuisine Information &
Flavour (spicy/mild), portion size, sauce details, temperature, cooking method, utensils, cultural norms, ingredient origin \\

Visual Information &
Multi-angle dish views; appearance as taste indicator \\

Navigation \& Interaction &
Bookmarking dishes; filtering by ingredient, flavour, portion size, cooking method \\

Social Context &
Cultural background understanding; low-pressure decision environment \\

\bottomrule
\end{tabular}
\end{table*}

\subsubsection{Ordering Efficiency Strategies}

Participants actively developed compensatory strategies to manage cognitive load under time
pressure. Seven participants first scanned section headers to isolate a preferred protein or
dish category. More than half searched for signal keywords such as \textit{traditional},
\textit{homemade}, or \textit{special} as proxies for safe or representative choices.
Several participants expressed a desire for secondary filter dimensions — by size, flavour, or
dietary composition — to accelerate selection. A small number preferred to bypass the menu
entirely and request direct recommendations from waitstaff, reflecting the degree to which
menu comprehension difficulties could erode confidence in independent decision-making.

\subsection{Mixed Media Evaluation and Design Expectations}

Following the immersive participatory design session, participants evaluated each medium on a
5-point Likert scale for \textit{likelihood of use} and \textit{perceived helpfulness for
understanding}. Table~\ref{tab:media_eval} presents the ratings alongside the key strength,
key limitation, and design expectations elicited for each medium — integrating what were
previously reported as separate findings (Sections~4.3 and~4.4 in the original manuscript)
into a single, structured reference.

\begin{table*}
\centering
\caption{Mixed Media Evaluation: Ratings, Strengths, Limitations, and Design Expectations (N=13)}
\label{tab:media_eval}

\begin{tabular}{p{1.5cm} p{1.2cm} p{1.8cm} p{2.6cm} p{2.6cm} p{3.7cm}}
\toprule
\textbf{Medium} &
\textbf{Likely (1--5)} &
\textbf{Helpfulness (1--5)} &
\textbf{Key Strength} &
\textbf{Key Limitation} &
\textbf{Design Expectations} \\
\midrule

Image &
4.63 &
-- &
Instant recognition; no learning required &
Static; authenticity concerns &
Interactive labels; multi-angle views \\

Video &
< Image &
> Image &
Shows cooking process and texture &
Time-consuming in decision context &
Controls, overlays, optional deep-dive layer \\

3D Model &
Moderate &
High &
Shows portion size and structure &
Requires high fidelity to be trusted &
Exploded view; scale reference; interaction \\

\bottomrule
\end{tabular}
\end{table*}

\subsubsection{Media Hierarchy Preferences}

Six participants expressed a preference for access to all three media types, with images
serving as the mandatory entry point. The preferred progression was:
\textbf{Image} (immediate, always visible) $\rightarrow$ \textbf{3D Model} (on demand,
for size and composition) $\rightarrow$ \textbf{Video} (optional, for deeper cultural
or preparation context). One participant proposed that video be hidden behind the dish
image by default, accessible only to users who actively choose to explore further.
Three participants preferred that video be excluded from the ordering flow altogether,
reserving it for post-dining or waiting-time engagement.

This preference for progressive, user-controlled disclosure — rather than simultaneous
presentation of all media — directly informed the layered information delivery framework
described in Section~5.

\section{Design implications based on user input }
\subsection{Optimizing mixed media Design for Variable Information Density and Accessibility}
During the interview about the three types of media, most of the participants mentioned that they think that video is informative, which is the most helpful media for them to learn more about the dishes. However, the likelihood of using it to decide which food to order is lower than in the other two media. In contrast, although people think they can hardly get more information about the dishes through images, it is most likely that they will use images to decide which dishes to order. 

From a mixed media design perspective, this suggested the importance of balancing information density and information usability.  For example, in a quick-decision-making scenario, designers should focus on the immediate clarification of the dishes by using images. In a context where participants think more detailed information is necessary,integrating videos that are concise yet informative could be more beneficial. A balanced mixed media approach that adapts to the context of use is essential. Therefore, designers can consider hybrid formats that combine the immediacy of images with clickable elements that expand to reveal more detailed video content. This allows users to quickly scan through options, while also having the opportunity to delve deeper if they desire more information.  This approach not only can cater to the preferences of different users, but also greatly improves user experience through different choices. 

\subsection{Prioritize the features}
The participants’ feedback also indicates prioritization of different media to present information about the dishes. Four participants suggested that when clicking the name of the dish, they hope that the images of the dishes can pop up immediately, followed by a 3D model which is further followed by videos. This indicates the method of layered information delivery. Considering that different types of information provided together may cause cognitive load, providing mixed media information in progressive disclosure can effectively solve this problem. From the most straightforward information provided by the image, to the dimensional information provided by the 3D models and finally to the deeper cultural information through video. 

The preference for choosing media is also influenced by the context of eating and the aim of eating in the restaurant. For example, some participants mentioned that it is more likely that they will watch the video or play with the 3D models if they eat individually in the restaurant compared to if they eat in a restaurant with families and friends. Other participants mentioned that the purpose of eating in the restaurant may be to chat, socialize, party, try new dishes, learn about new culture. In some circumstances, interaction between people can be more important than understanding the menu and ordering food. Therefore, different restaurants' contexts can closely influence the type of media people select to understand the dishes and order the food. 

Considering both aspects, the design strategy that users should be allowed the customization of the media and information can be developed. Many participants mentioned that they would like different media to be available but not forced to use them. Providing an interface with buttons for different media allows users more freedom to choose the information they want, which can provide personalized experiences and address needs in various contexts.

\subsection{Category system}
Our findings reveal a significant need for a more flexible and user-centric approach to menu categorization. The participants expressed various preferences for organizing the options for the cuisine, suggesting that a one-size-fits-all approach is insufficient for cultural newcomers navigating foreign menus. During the interview, participants showed interest in browsing cuisines using cooking techniques, ingredient types, and portion sizes, among other criteria. 10 out of 13 participants expressed a desire for additional categorization methods beyond the traditional menu structure. Customizable categories were perceived to potentially improve the efficiency of the order process.

Implement a dynamic and multifaceted categorization system within the MR interface that allows users to reorganize menu items according to their preferred criteria. This system should offer Multiple Categorization Options (e.g., cooking techniques, main ingredients, portion sizes, dietary restrictions). This feature could enable cultural newcomers to explore menus in ways that align with their thought processes and preferences and reduce the time spent deciphering unfamiliar menu structures, potentially leading to faster and more confident ordering decisions.

\section{Discussion}
Based on the feedback of the participants on current digital tools, different types of media, and their expectation about the three types of media and how they can be used to understand the menu and order food, we generated some design strategies to guide the design of the MR tool to improve the dining experience of cultural newcomers in restaurants. The next step would be implementing the MR virtual companion for dining based on the design insights put forward, and testing its effectiveness. Considering the close relationship between food and culture, more research is also needed to determine whether ease of dining can facilitate cultural perception and thus contribute to cultural adaptation of cultural newcomers. 

However, current research has several limitations. First, the current participants are mainly from China,  so there may be some bias in the data collected. The second issue is that the number of participants in the interview is not that large, which makes the result of the investigation not as inclusive as possible. In addition, this research mainly focuses on the influence of the menu in another language on food ordering in the restaurant and ignores other elements, such as the interaction between customers and waiters/waitresses and
the attitude of waiters/waitresses. 

\section{Conclusion}
As the world becomes increasingly interconnected, more cultural newcomers (CNs) face significant challenges in navigating unfamiliar culinary landscapes. Our study has illuminated the complex interplay of linguistic, cultural, and cognitive barriers that CNs encounter when ordering food in foreign restaurants. By exploring the potential of mixed media integration in mixed reality environments, the research uncovered potential avenues for enhancing menu understanding and order decision making.
Our findings reveal the limitation of current digital tools in providing a comprehensive and efficient solution for menu comprehension and food ordering. The need to switch between multiple applications and the lack of contextual information highlight the limitations of existing technologies. In response, our participants expressed a clear desire for more intuitive, mixed media-rich solutions that can provide immediate, visually oriented information about dishes.

The exploration of various media types - images, videos, and 3D models and their possible integration format in an MR setting - has demonstrated the potential for a more immersive and informative dining experience. These insights point towards a future where MR technologies can play a crucial role in bridging cultural and linguistic gaps in everyday scenarios like dining. By integrating mixed media elements in a thoughtful, user-centric manner, we can create tools that not only facilitate menu comprehension, but also enhance cultural learning and adaptation.

In conclusion, our research underscores the potential of MR and mixed media integration to address the daily challenges faced by cultural newcomers. Research can also benefit the implementation of mixed media in MR to facilitate causal learning and cultural adaptability.

\begin{acks}
We thank all participants who contributed their time and insights to this study. We also acknowledge the support of our research group and colleagues who provided valuable feedback during the development of this work.
\end{acks}

\bibliographystyle{ACM-Reference-Format}
\bibliography{references}

\end{document}